\documentclass[aps,prd,showpacs,showkeys,preprint,amsmath,amssymb]{revtex4}
\usepackage[english]{babel}
\usepackage{epsfig,float,subfigure}
\makeatletter

\newcommand{\beq}{\begin{equation}}
\newcommand{\eeq}{\end{equation}}
\newcommand{\bea}{\begin{eqnarray}}
\newcommand{\eea}{\end{eqnarray}}

\usepackage{amssymb,amsmath}

\begin{document}
\title{Exact solutions for a Maxwell - Kalb-Ramond action with dilaton:
localization of massless and massive modes in a sine-Gordon brane-world}
\author{H. R. Christiansen, M. S. Cunha, M. O. Tahim}
\affiliation {Grupo de F\'{\i}sica Te\'orica, State University of Ceara (UECE) -  Av. Paranjana,
1700, CEP 60740-903, Fortaleza, Cear\'a, Brazil }
%

\begin{abstract}
We analytically find the exact propagation modes of the electromagnetic {and} the
Kalb-Ramond fields \textit{together} in a five-dimensional curved space-time. The existence
and {localization of gauge particles into our four-dimensional world (4D)}
is studied in detail on a brane-world scenario
in which two gauge fields interact  with a {dilaton} and a gravitational background.
The coupling to the dilaton is different in each case causing the splitting
between both gauge spectra.
The gauge field zero-modes and an infinite tower of Kaluza-Klein massive
states are \textit{analytically} obtained. Relevant conditions on the dilaton coupling
constant are found in order to identify with precision \textit{every finite tensor and vector
eigenstate in the theory}. An \textit{exact quantization condition on the whole mass spectrum},
depending on the dilaton coupling constant and the bulk Planck mass, is inherited
from the extra-dimension. {This allows finding an exact rule to prevent tachyons
in the theory and, by the same token, predicting a possible Kalb-Ramond tensor zero-mode
in 4D world. We also show that KK massive-modes contributions onto 4D
physics are strongly suppressed}. \end{abstract}

11.10.Kk., 04.50.-h. \hfill \texttt{To appear in Phys.~Rev.~D, 2010}


\maketitle
\section{Introduction}

Together with the mass question, one of the most difficult issues to
understand in the Standard Model is how the electroweak scale can be
perturbatively stable at seventeen orders of magnitude below
the Planck scale. In other words,  whether there is a way to bring
the quantum gravity scale below $\sim$1 TeV.

Recent proposals involving high-dimensional models and brane-world
scenarios raise possible solutions to such a hierarchy problem of
gauge couplings \cite{RS}.
Ten-dimensional superstring theory allows these kind of frameworks
to get embedded within a theory of everything. Since extra-dimensional
theories present more
degrees of freedom than four-dimensional (4D) ones, they provide a
richer framework to
approach physical phenomena. The problem is that the parameter space
to be covered is much wider and calculations in higher dimensional
gravity are technically very difficult.

The main goal of high-dimensional
theories being the inclusion of gravity together with the
Standard Model interactions introduces
nontrivial changes in all sectors, so gauge forces
have to be proven to exist as needed in the ordinary space slice.
In the gauge sector the analysis of localization becomes crucial to have
a 4D effective result with massless photons.
Modeling our universe as a domain wall has been
motivated by the D-brane solutions of string theory.
Gauge fields are deposited on D-branes from open strings ending on them.
However, domain walls can be constructed in field theoretic frameworks
with no direct possibility to localize gauge fields. As a matter of fact,
in 5D we shall need not only gravitational fields but also a dilaton
field to properly accomplish
this task \cite{keha-tamva, youm1}. Since simple domain walls cannot
hold gauge fields in \cite{pomarol},
this scalar will show to be essential to guarantee the existence
of U(1) gauge fields localized in the four-dimensional world.

Current interest in theories with bosonic fields of different spin
arise as a result of their existence on conformally flat spaces of
any dimensionality ($D\geq 4$) such as anti-de Sitter geometries.
The role of tensor bosons in the context of AdS/CFT correspondence,
is particularly important when coupled to gravity \cite{spin ads-cft}.
String theory is so far mainly based on its low spin excitations and
their low energy interactions.
The low energy 4D effective field theory action of string theory \cite{polchinski}
contains two massless fields: a second rank antisymmetric tensor
from the (Neveu-Schwarz) NS-NS sector of the underlying string theory known as
Kalb-Ramond field \cite{KR} and a scalar field called dilaton.
The Kalb-Ramond field, provided the lagrangian is invariant under a second gauge
transformation, 
has three free components. However, it is known that classically a
free antisymmetric tensor field is dual to a scalar which involves
just one degree of freedom.
Indeed, the third rank field-strength corresponding to the the Kalb-Ramond field is dual to a
pseudo-scalar in 4D so-called axion. Consequences of the presence of the axion
in a curved space-time on some physical phenomena have been investigated in view of possible
indirect evidences of string theory at low energies \cite{axion}.
%

Here, we will focus on {both} Maxwell and Kalb-Ramond gauge fields together,
in a warped five-dimensional bulk with a dilaton and a
brane defect that mimics the ordinary world. This thick brane configuration
results from a field action with a (deformed) sine-Gordon potential that depends
on two scalars. One of these is a kink representing the membrane itself and the other
the dilaton in a field theoretic scenario. Metric, dilaton
and brane configurations are geometrically consistent solutions of the world action.
On such an interesting background we shall \textit{analytically obtain
all the eigenstates for both gauge fields, viz. zero-modes, massive modes and tachyons
building both whole spectra}.

{Regarding the propagating modes of the Kalb-Ramond field, so far only
simple and qualitative calculations have been obtained in this context. In the present paper,
we perform a systematic search by means of an analytical approach. Furthermore,
we also include the electromagnetic field together, in order to exhibit their actual differences.
We thus show that vector and tensor spectra are analogous but not equal as a consequence
of their different coupling to the dilaton. In both cases, the full tower of gauge massive modes,
so far invisible in ordinary world, are proven to be strongly suppressed in 4D. We also predict
that a localized zero-mode is not only possible for the 5D electromagnetic field
but also, and simultaneously, for a 5D Kalb-Ramond field, provided the dilaton coupling constant is above a certain value.
This value is precisely that needed to exclude tachyons in the theory.
All of this amounts to show that the model presented here
is an interesting arena to discuss extra-dimensional physics and that our ordinary 4D world
seems to be compatible with a higher dimensional universe, apparently of a 4D stringy brane nature.}

{}

The paper is organized as follows. In the next Section, we present the geometrical
background. Along with a conveniently warped metric, this includes a
sine-Gordon potential depending on the brane and the dilaton fields. In previous treatments
a Higgs like potential has been preferred, so we shall briefly address this case as a warm up.
In Section \ref{sect action} we introduce the action for the gauge fields in the bulk,
Maxwell and Kalb-Ramond, coupled to a warped gravity and
dilaton background. We next derive the bulk equations of motion and separate
the extra coordinate from the four-dimensional world coordinates. Then, in Section
\ref{sect spectrum}, we find a condition for localization and transform the equations
of motion for the gauge fields into a couple of Schrodinger like equations. In the following sections
we analytically discuss the Maxwell and Kalb-Ramond spectra as a function of the dilaton
coupling constant. We particularly emphasize the Sturm-Liouville nature of the differential equations
resulting from the mapping of the original problem, and show significant consequences on the space of
solutions.
In Section \ref{sect complete} we obtain exact expressions for the
full spectrum of eigenstates of the problem. Analyticity constraints on massive modes,
zero-modes and tachyons are discussed.
Kaluza-Klein eigenstates are fully exhibited and we show that massive modes are
strongly suppressed on the brane so that ordinary four-dimensional gauge
interactions are not significantly  modified in the present set up.
Finally, in Section \ref{sect conclusion} we draw our conclusions.

\section{The space-time background: gravitational warping and dilaton \label{sect background}}

We start our analysis by studying the appropriate  space-time framework
for the description of a consistent gauge theory.
In this Section we will show how one proceeds with both vector and tensor gauge
fields in order to look for zero-modes that can be localized in a four-dimensional membrane
embedded in a five-dimensional space-time.
Here, the extra dimension is
not necessarily an orbifold and will be assumed infinite. Furthermore, the brane is not included
in the model as a static external source but it is dynamically obtained as a solution to the
Einstein equations
for gravity coupled to (two) scalar fields. One of these scalars is for creating
a domain wall defect (a thick brane) while the other is the dilaton field.

It is worth noting that either gravitational and fermionic massless quanta can be trapped
in four-dimensional (4D) domain walls lying within a five-dimensional
(5D) bulk with some {AdS} like metric \cite{keha-tamva, massless modes}.
On the other hand, it is known that vector gauge bosons in these kind of scenarios
are not localizable unless the coupling constant is dynamically modified.
The reason is that gauge field theory is conformal
\cite{dvali shifman} so that all the information coming
from the metric warping factors automatically drops out resulting in a non-normalizable zero-mode.
Fortunately, the coupling of the dilaton to
the gauge field in the kinetic term modifies the rescaling properties
allowing for the localization of the vector zero-mode \cite{keha-tamva} and
tensor zero-mode \cite{prd tahim} respectively (see also \cite{yaoum2}).

Following this approach and motivated by low-energy string theories, here we will discuss
both Maxwell and Kalb-Ramond gauge fields together, coupled to a
gravitational background and a consistent dilaton configuration.
In what follows, we shall obtain close expressions
for the gauge modes in the 5D space-time. This will be in order to
analytically discuss the phenomenological consequences, not only of zero-modes
but also of massive states.

First, it is necessary to obtain a solution to the equations
of motion of the gravitational field for a
potential functional depending on both the dilaton and membrane field variables \cite{keha-tamva}.
We therefore analyze the following action for two real
scalar fields
\begin{equation}
S_B=\int d^{4}x\ dy
\sqrt{-\det G_{MN}}\ [2M^{3}R-\frac{1}{2}(\partial\Phi)^{2}-\frac{1}{2}%
(\partial\Pi)^{2}-\mathcal{V}(\Phi,\Pi)],
\label{action bounce}
\end{equation}
where $M$ is the Planck constant in 5D, and $R$ is the Ricci scalar. 
The solution for $\Phi$ is the membrane kinking on our 4D-world.
The corresponding field solution
for $\Pi$ will be the dilaton configuration consistent with the metric and the kink.
As usual we adopt latin capitals on the bulk and greek lower case letters on 4D.

We next shall assume some ansatz for the space-time metric
\begin{equation}
ds^{2}=e^{2\Lambda(y)}\eta_{\mu\nu}dx^{\mu}dx^{\nu}+e^{2\Sigma(y)}dy^{2},
\label{warpedmetric}
\end{equation}
where $\Lambda$ and $\Sigma$ are warp functions that depend just on
the extra (fifth) coordinate, and diag$(\eta)=(-1, 1, 1, 1)$. The
equations of motion for action (\ref{action bounce})
are
\bea &&\frac{1}{2}(\Phi^{\prime})^{2}+\frac{1}{2}(\Pi^{\prime})^{2}-
e^{2\Sigma(y)}\mathcal{V}(\Phi,\Pi)=24M^{3}(\Lambda^{\prime})^{2},\label{motion1}\\
&&\frac{1}{2}(\Phi^{\prime})^{2}+\frac{1}{2}(\Pi^{\prime})^{2}+e^{2\Sigma(y)}\mathcal{V}(\Phi,\Pi)=
-12M^{3}\Lambda^{\prime\prime}-24M^{3}(\Lambda^{\prime})^{2}+12M^{3}\Lambda^{\prime}\Sigma^{\prime},
\nonumber\eea and  \bea
\Phi^{\prime\prime}+(4\Lambda^{\prime}-\Sigma^{\prime})\Phi^{\prime}=
e^{2\Sigma}\ \frac{\partial\mathcal{V}}
{\partial\Phi},\nonumber\\
\Pi^{\prime\prime}+(4\Lambda^{\prime}-\Sigma^{\prime})\Pi^{\prime}=
e^{2\Sigma}\ \frac{\partial\mathcal{V}}{\partial\Pi}.
\label{motion2} \eea where the prime means derivative with respect
to $x^5=y$ (note the correction in eq.(\ref{motion2})
with respect to Ref.\cite{keha-tamva}).

In order to solve this system we use a supergravity motivated \cite{supergravity}
potential functional $\mathcal{W}(\Phi)$. This so-called superpotential is applicable to
non-supersymmetric domain walls
\cite{superpotential} as the present one, and is defined by \beq\Phi^{\prime}=\frac{d
\mathcal{W}}{d\Phi}.\label{W}\eeq
In the absence of gravity, for a double-well potential of the Higgs type
$V(\Phi)=\frac{\delta}4 (\Phi^2-v^2)^2$, the simplest possible static
membrane configuration dependent on the fifth coordinate is a bounce
\beq \Phi(y)=v\tanh(\kappa y)\label{bounce}\eeq  \cite{keha-tamva}, consistent with the
superpotential
\begin{equation}
\mathcal{W}(\Phi)=v\kappa\Phi(1-\frac{\Phi^{2}}{3v^{2}}),
\end{equation}
where $\kappa^2=\delta v^2/2$.

Putting into scene the metric ansatz (\ref{warpedmetric}), the
dilaton field $\Pi(y)$, and taking into account the equations of motion, the
necessary potential consistent with the bounce can be written as
\begin{equation}
\mathcal{V}(\Phi,\Pi)=\exp{({\Pi}/{\sqrt{12M^{3}}})}\cdot
\left(\frac{1}{2}(\frac{d\mathcal{W}}{d\Phi})^{2}-\frac{5}{32M^{3}}\mathcal{W}(\Phi)^{2}\right).
\label{potential}
\end{equation}
Now, one obtains the following solution to the   equations of motion,
eq.(\ref{motion1}) and eq.(\ref{motion2}),  consistent with the bounce
configuration eq.(\ref{bounce}) and the potential eq.(\ref{potential})
\beq \Lambda(y)=4\ \Sigma(y)=\frac{-\Pi(y)}{\sqrt{3M^3}}=-\beta
(\ln\cosh^2(\kappa y)+\frac 1 2 \tanh^2(\kappa y)),\label{solution}\eeq where
$\beta=v^2/36M^3$.

\begin{figure}
\includegraphics{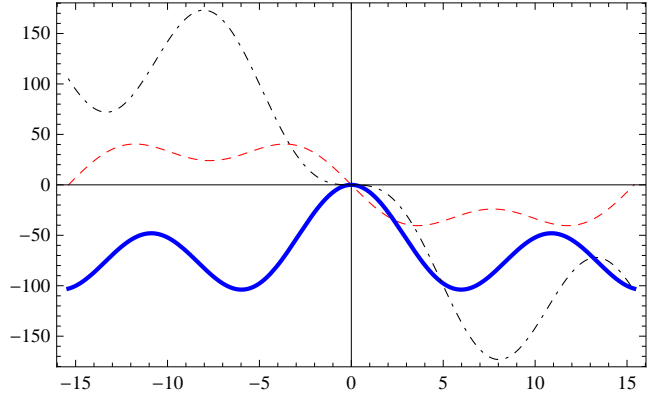}
\caption {\label{fig1(potphi)} Modified background potential $\mathcal{{V}}(\Phi)$
(eq.\ref{newpotential phi}) for
different values of $a=1/6M^3b^2$: $a=1$ (dashed red line), $a=2$ (solid blue line), $a=3$ (dash-doted black line).}
\end{figure}

The possibility of considering a stack of traveling branes of this kind, or even colliding branes,
can be addressed by means of a sine-Gordon potential, which in the absence of gravity
is very well-known and has the form
\beq V(\Phi)=\frac 1 {b^2}\ (1-\cos (b\,\Phi)).\eeq
Here, $b$ is a free parameter which shall be related to
the asymptotic curvature when the theory includes gravity.
With this potential, new solutions connecting separate vacua
are possible. One-soliton solutions for this potential read
 \beq
\Phi(y)=\frac 4 b \ \arctan e^y\label{newbounce}\eeq
(we shall assume it static for simplicity).
Now, in a gravitational background it is still possible that this bounce represents
a legitimate brane-world. Considering the gravitational ansatz (\ref{warpedmetric}), the
equations of motion (\ref{motion1}) and (\ref{motion2}) are
compatible with this solution provided we somehow modify the
potential functional. Taking into account eq.(\ref{W}) we now obtain the
superpotential
\begin{equation}
\mathcal{W}(\Phi)=-\frac 4{b^2}\cos(\frac {b}2\,\Phi).
\end{equation}
From  (\ref{potential}) we get the following
 background  potential
\begin{equation}
\mathcal{{V}}(\Phi, \Pi)=-\exp{({\Pi}/{\sqrt{12M^{3}}})} \cdot
\left(\frac 4{b^2}\sin^2(\frac b 2\,\Phi)+\frac{5}{2M^3 b^4}\cos^{2}\frac b 2\,\Phi\right)
\label{newpotential}
\end{equation}
which can also be written as a function of $\Phi$ by means of

\begin{equation}
\mathcal{{V}}(\Phi)=-\frac 4 {b^2}\ (\sin \frac b 2 \Phi)^{1/6M^3{b^2}}\
\left(1+(\frac{5}{8M^3 b^2}-1)\cos^{2}\frac b 2\,\Phi\right),
\label{newpotential phi}
\end{equation}
both exhibiting a highly nontrivial dependence on the scalar fields (see Fig. \ref{fig1(potphi)}).
\bigskip

By writing the hamiltonian \`a la Bogomol'nyi, it can be
deduced that $\mathcal{{V}}$ is consistent with the following relations among the
warping, dilaton and superpotential  \beq
\Pi=-\sqrt{3M^3}\Lambda, \ \ \Sigma=\Lambda/4, \ \ \Lambda^\prime=-
\mathcal{W}/12M^3. \label{newsol}\eeq
Solving the equations of motion, the explicit dependence of the dilaton field
on the extra dimension is given by
\beq \Pi(y)= \frac 1 {\sqrt{3M^3} b^2} \ln\cosh y\label{newsolution},\eeq
from which we can readily obtain all the other field distributions
of the background configuration.

As usual with dilaton configurations related to D-brane solutions, these functions
are singular as $|y|\rightarrow\infty$.
However, since the metric vanishes exponentially and both dilaton and warp factors
operate by means of a negative
exponential coupling, at the end of the day the model is free of divergences.
Indeed, as we will see in the next Section, the effective action remains finite
precisely thanks to the dilaton configuration.
Note that the curvature of the metric with dilaton, as given by
the Ricci scalar, now reads
\begin{equation}
R(y)=-[8\Lambda''(y)+18(\Lambda'(y))^{2}]\exp\left({\frac{\Pi(y)}{2\sqrt{3M^{3}}}}\right),
\end{equation}
which for the background just obtained results in
\begin{equation}
R(y)=16\ a (\cosh y)^{a-2}  (1-\frac 9 2 \sinh^2y),
\label{ricci}
\end{equation}
where $a=1/6M^3b^2$. Since the dilaton contribution amounts to a
redefinition of the effective four-dimensional Planck scale,
eq.(\ref{ricci}) exhibits a negative growing behavior far from the membrane.
However, this problem 
disappears when we lift the metric solution up to $D=6$
\cite{d=6 metric}, where the dilaton represents the radius of
the new extra dimension as  happens in type II string theory with
D4 branes when lifted into D=11 supergravity. As explained in \cite{keha-tamva},
in six dimensions $ds^2_6=e^{3A(y)/2}(-dt^2+dx_1^2+dx_2^2+dx_3^2+dz^2)+dy^2$, where $z$
parametrize an extra $S^1$ direction, and this metric results everywhere regular
for the solutions given above.

%
%
%

Studying as well the fluctuations of the metric about the above configuration,
it is possible to see that this model supports a massless zero-mode of the gravitational
field localized on the membrane even in the dilaton background. In order to prove the
stability of the background solution,
we would have to show that there are no negative mass solutions to the equations of motion
of a perturbation $h_{\mu\nu}$ of the metric.
Actually, a graviton massive spectrum appears starting from zero and presenting no gap.
This can be easily seen by means of a supersymmetric type expression
of the Schrodinger type operator which results after an appropriate change of variables
and decomposition of the graviton field (see \cite{keha-tamva, gremm} for details).
The issue of the coupling of these massive modes to the brane has been analyzed in detail
in \cite{csaki al}.

\section{Vector and tensor gauge fields in a warped space with dilaton \label{sect action}}

Now let us consider the system of five-dimensional electromagnetic $A_{N}$
and Kalb-Ramond $B_{NP}$ gauge
fields coupled to the dilaton in a warped space-time. We will adopt
the following 5D action

\begin{equation}
S_g = \int dy\,d^4x\,\sqrt{-\det G_{AB}}\ e^{-\frac\lambda
2 \Pi }\ \left\{\dfrac1{12}\,
e^{-\frac\lambda 2 \Pi }H_{MNP}H^{MNP}-\dfrac14\,F_{MN}F^{MN} \right\}
\label{action gauge}
\end{equation}
where $H_{MNP}=\partial_{[M}B_{NP]}$ and $F_{MN}=\partial_{[M}A_{N]}$.

Assuming that the gauge field energy density should not
strongly modify the geometrical background, we can study the
behavior of propagating modes in the background of the topological
configuration  studied in the last Section.
In this respect, for example, Das et al. \cite{KR RS}
derived an exact solution for the
metric of a Randall-Sundrum (RS) approach  with a Kalb-Ramond term (with no dilaton) and showed
a negligible deviation from the pure RS solution without gauge fields.
The new metric depends on
the energy density of the Kalb-Ramond field and goes smoothly to the RS solution
 in the limit of Kalb-Ramond energy density tending to zero. This
scenario solves the hierarchy problem  not just for the
orbifold radius predicted by RS but for any value greater than the
RS value. However, the important point here is that the Kalb-Ramond energy
density is insignificant, amounting to $\sim 10^{-62}$.
Indeed, this value matches remarkably well with the Kalb-Ramond
energy density on the visible brane, calculated from the solution of
the Kalb-Ramond field in a Randall Sundrum brane-world in a
previous work \cite{previous KR RS}. 
%
In general, most of the attempts to stabilize 5D brane worlds by means of a scalar field in
the bulk do not take into account the back-reaction of the scalar field on the
background metric \cite{keha-tamva, scalar RS, gremm, csaki al} and those in order to compute the
scalar back-reaction on the metric
were unsuccessful except in a few special cases \cite{scalar backreaction}.

The dilaton $\Pi(y)$ couples exponentially with the kinetic terms of both
gauge fields $B_{MN}$ and $A_{M}$ \cite{dilaton coupling}. This is related
to the fact that the combination $\sqrt G\, \exp\Pi$ can be interpreted as a change
in the integration measure. Still, different coupling constants ($\lambda$
and $\lambda/2$) are assigned to the 3-rank tensor $H_{MNP}$
and the electromagnetic field-strength $F_{MN}$ respectively  \cite{mayr}.
This will be the origin of two different
equations of motion for either gauge field as we now show.

It is easy to see that action (\ref{action gauge}) is invariant under
gauge transformations $\delta B_{MN}=\partial_{[M}\Omega_{N]},
\ \delta A_{M}=\partial_{M}\omega$.
In five dimensions, the equations of motion for $B_{MN}$ and $A_M$ are given by
\begin{eqnarray}
\frac 1{\sqrt{-G}}\,\partial_M (G^{MR}G^{NP}F_{RP} \sqrt{-G} e^{-\frac \lambda 2\Pi(y)})=0\nonumber\\
\frac 1{\sqrt{-G}}\,\partial_M (G^{MR}G^{NS}G^{PQ} H_{RSQ}\sqrt{-G}
e^{-\lambda  \Pi(y)})=0 \label{more motion}
\end{eqnarray}
where diag $G_{MN}=(e^{2\Lambda}\eta_{\mu\nu}, e^{2\Sigma})$.

In order to solve these equations, we adopt the following gauge choices \beq
A^{5}=0,\ \partial_{\mu}A^{\mu}=0,\ B^{\mu 5}=0,\
\partial_{\mu}B^{\mu\nu}=0.\label{gauge choice}\eeq
Next, we write down the equations in contravariant components
({$T^{MNP\dots}= G^{MR}G^{NS}G^{PQ}\dots T_{RSQ\dots}$}) and separate the fifth
from the other coordinates as follows
\beq A^{\mu}(x, y)=a^{\mu}(x)u(y), \ \ \ \ B^{\mu\nu}(x, y)=b^{\mu\nu}(x)w(y).\label{decomposition}\eeq

Now, from eq.(\ref{more motion}) we just get
\bea
&& [\Box +\frac 1{u\ f} \partial_5(f\partial^5 u)]\ a^{\mu}=0\\
&& [\Box +\frac 1{w\ g} \partial_5(g\partial^5 w)]\ b^{\mu\nu}=0.
\eea
Note that the metric deforms the otherwise trivial solutions of this system
of equations through
the warping functions and the dilaton field by means of the factors
$$f(y)=\exp[4\Lambda+\Sigma-\lambda\Pi/2],\ \ \ \ g(y)=f\exp[-\lambda\Pi/2]$$
acting on $u(y)$ and $w(y)$ respectively.

We may reduce the space of solutions to a rather special case. If we assume
$u(y)=u_{0}$ and $w(y)=w_{0}$, normalizable zero-modes for the gauge fields
can be obtained when $u_0$ and $w_0$ are
(nonzero) constants. On the other hand, Kaluza-Klein modes result
from the solution of the general case
\beq \partial_5(f\partial^5 u)=-m_A^2 f u,\ \
\partial_5(g\partial^5 w)=-m_B^2 g w
\eeq
where $m_A^2,\ m_B^2$ are arbitrary constants representing the 4D squared bosons masses
of vector and tensor gauge fields respectively. It means that $a^{\mu}=e^{ipx}$
with $\eta^{\mu\nu}p_\nu p_\mu=p^2=-m_A^2$, and $b^{\mu\nu}=e^{ikx}$ with $k^2=-m_B^2$.
Below, we will show that these constants are indeed quantized eingenvalues of a
Schrodinger-like equation. Furthermore, we will show that $u(y)=u_{0}$ and $w(y)=w_{0}$
are not just some special case but the unique zero-modes of the theory.

Explicitly, the most general $y$-dependent equations
of motion from action (\ref{action gauge}) for the modified sine-Gordon potential
(\ref{newpotential phi}) read

\beq
 u''(y)+ a (1-2c_1)\tanh y\ u'(y) + m_A^2\cosh^{-a}y\ u(y)=0\label{massive modes}\eeq
\beq
w''(y)+ a (1-2c_2)\tanh y\ w'(y) + m_B^2\cosh^{-a}y\ w(y)=0,\label{massive modes 2}
\eeq
where $c_1=(17+2\lambda\sqrt{3M^3})/4$ and $c_2=(17+4\lambda\sqrt{3M^3})/4$.
We can see that the different dilaton coupling to the Maxwell and Kalb-Ramond fields is
responsible for the different massive modes of vector and tensor bosons. Zero modes, however,
are identical for both fields in this model. Nevertheless, as we will show later on, they are
not necessarily localizable together.
\section{Analysis of the gauge-bosons spectra \label{sect spectrum}}

In this section we will discuss the existence of gauge field solutions to the model and their
localization on the membrane.
We can probe the localization of the gauge field modes by verifying that the corresponding
action is finite. Note that from eq.(\ref{gauge choice}) and eq.(\ref{decomposition}) it follows
$H^{\mu\nu\rho}=h^{\mu\nu\rho}w(y)$ and $F^{\mu\nu}=f^{\mu\nu}u(y)$, so that
\bea
S_g [\rm sol.] = && \int dy\ u^{2}(y)e^{4\Lambda(y)+\Sigma(y)-\lambda\Pi(y)/2}\int
d^{4}x\ \frac 1 4 f_{\mu\nu}f^{\mu\nu}-\nonumber\\
&& \int dy\ \frac 1 {12}w^{2}(y)e^{4\Lambda(y)+\Sigma(y)-\lambda\Pi(y)}\int
d^{4}x\ h_{\mu\nu\alpha}h^{\mu\nu\alpha}.\label{finiteaction}
\eea
For constant $u(y)=u_{0}$ and $w(y)=w_{0}$, which satisfy the equations (\ref{massive modes})
 and (\ref{massive modes 2}) for zero mass, this integral can be analytically
proved to be finite. This shows
that zero-modes associated with the Maxwell and Kalb-Ramond fields in the
dilaton background can be localized on the four-dimensional space-time of a kink. Using
the solutions found in eq.(\ref{newbounce}) and equations thereafter, we have

\beq \sim\int dy\ e^{4\Lambda(y)+\frac 1 4\Lambda(y)+\frac{\lambda}2\sqrt{3M^3}\Lambda(y)}
\eeq
which according to the solution
\beq
\Lambda(y) = 2a\ln \rm sech y \eeq
(c.f. eq.(\ref{newsol}) and eq.(\ref{newsolution}))
and the definition of constants $a$ and $c_1$, results in
\beq
\int dy\ e^{c_1\Lambda} =
\int dy\ \cosh^{-2ac_1} y <\infty \label{localization}
\eeq
provided $c_1>0$, namely $\lambda> -\frac{17}{2\sqrt{3M^3}}=\lambda_0$.
If  $c_2>0$, then $\lambda>\lambda_0/2$ and there should be also localized
zero-modes for the Kalb-Ramon field for any possible value of $\lambda$ in this interval.


Regarding Kaluza-Klein modes, eq.(\ref{massive modes}) and (\ref{massive modes 2}) could be
numerically solved to have an idea about some particular cases. However,
since an ordinary differential equation can always be put into its normal form,
we can have a deeper insight on the whole problem. For this, let us then make the following
transformation in eq.(\ref{massive modes}) (see e.g. \cite{RS, keha-tamva})
\beq  u(y)=e^{-\alpha \Lambda/2}U(z),\ \ \ \frac{dz}{dy}= e^{-\beta \Lambda}\label{change variables} \eeq
where $\alpha$ and $\beta$ are arbitrary parameters.
We now set $\alpha=c_1-1/4$ and $\beta=-1/4$ in order to eliminate the first derivative
term in $U$ and to have a pure mass term.
This brings eq.(\ref{massive modes}) into a simplest case of the Sturm-Liouville
equation (with respect to a weight function 1). This is precisely a
Schrodinger-like equation in the variable $z$
\beq \left[-\frac{d^2}{dz^2}+ \mathfrak{V}(z)\right]U(z)=m_A^2U(z),\label{schrodinger}\eeq
where $\mathfrak{V}(z)=e^{-\Lambda/2}(\frac{\alpha}2\Lambda''-\gamma\Lambda'^2)$ and
$\gamma=\frac 1 4\alpha(\frac 1 2-\alpha)$.
If we now choose $a=2$
we arrive at the following expression for such analog non-relativistic potential
\beq \mathfrak{V}(z)=-2\alpha\ \left[1-(2\alpha-1)\tan^2z\right]
\label{potential tan2} \eeq
which is a fully tractable function as we will see in what follows.
Eq.(\ref{massive modes 2}) can be dealt with identically and  we let it be
considered in the next Sections.

\subsection{An exact subset of eigenstates \label{subsect exact}}

Interestingly, the Schrodinger equation we have just identified has been recently
discussed in the literature in a very different context. In \cite{taseli} exact
bound states have been analytically found for  arbitrary $\alpha\geq 1/2$.
Actually, bound states should be also possible for $\alpha\leq0$ (besides localization
of zero-modes precludes the region $\alpha\leq -1/4$) but this will be discussed in
the next Section. Since $\mathfrak{V}(z)$ is defined within
$-\pi/2-2\pi k < z <\pi/2+2\pi k, \ k\in N$, bound eigenstates of eq.(\ref{schrodinger})
have to vanish at the endpoints, i.e. $U(z)|_{z=\pm(\pi/2+2\pi k)}=0$. The reflection symmetry of
the potential indicates that the eigenfunctions are exactly classified into two classes, symmetric
and antisymmetric. These bound states are of course normalizable. However, in order to study
normalization in the real bulk, we have to anti-transform the solution back to the $u(y)$ space.
It can be seen that there the only situation could arise in the case of positive $\alpha$ for
large values of $y$; however, these correspond to values of $z$ close to $\pi/2$ (and multiples)
which are forbidden places for such a potential.

The symmetric eigenstates of eq.(\ref{schrodinger}) can be generally written in
an exact way in terms of Gauss
hypergeometric functions
\smallskip
\begin{figure}[h]
 \centering
\includegraphics[width=7.5cm,height=5.5cm]{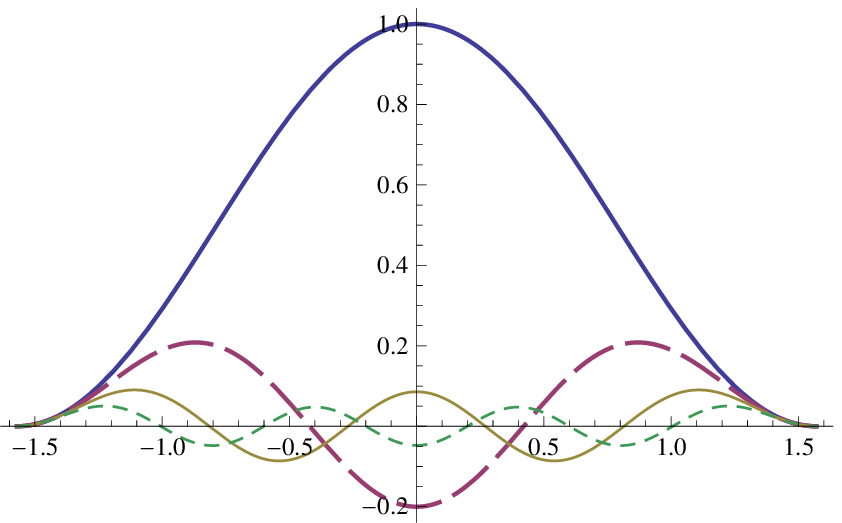}
\caption{\label{figU} Symmetric $U_{2n}(z)$ functions for $\alpha = 1$
and $n=0$ (blue solid line), $n=1$ (red dashed line), $n=2$ (yellow solid line), $n=3$ (green short-dashed line).}
\includegraphics[width=7.5cm,height=5.5cm]{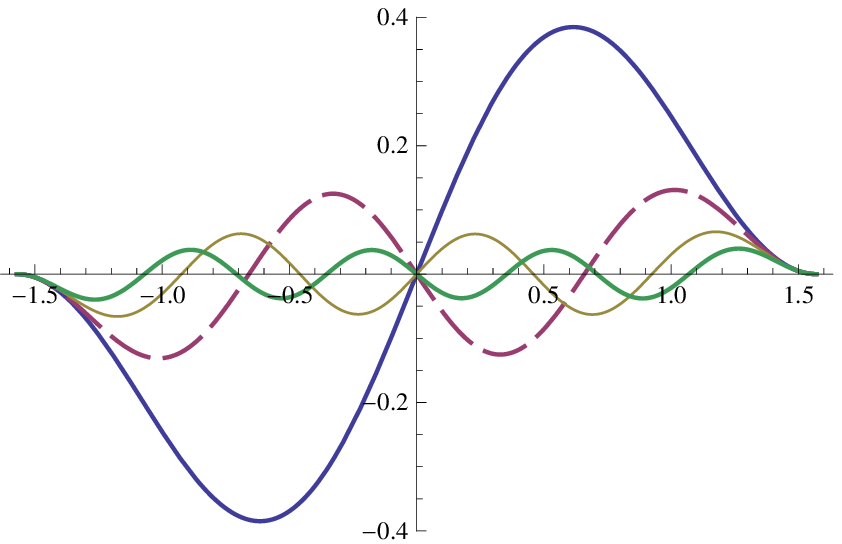}
\caption{\label{figUa} Antisymmetric $U_{2n+1}(z)$ functions for $\alpha = 1$ and $n=0$ (blue solid line), $n=1$ (red dashed line), $n=2$ (yellow thin-solid line), $n=3$ (green solid line).}
\end{figure}
\beq U_{2n}(z; \alpha)= A_{2n}(\alpha)(\cos z)^{2\alpha}\,\,
{}_2F_1\left(-n, 2\alpha+n, 2\alpha+\frac 1 2; \cos^2z\right),\label{symmetric states U}\eeq
where
$${}_2F_1\left(a, b, c; d\right)= 1+\frac{a\ b\ d}{c\ 1 ! }+\frac{a(a+1)\ b(b+1) d^2}{c(c+1)\ 2 !}+\dots,$$
and $A_{2n}(\alpha)$ is a normalization constant.

In the case of anti-symmetric eigenstates, the exact solutions read
\beq U_{2n+1}(z; \alpha)= A_{2n+1}(\alpha)\sin z\ (\cos z)^{2\alpha}\,\,
{}_2F_1\left(-n, 2\alpha+n+1, 2\alpha+\frac 1 2; \cos^2z\right),
\label{antisymmetric states U}\eeq
for $n=0, 1, 2, 3, \dots$ (see Fig. \ref{figU} and Fig. \ref{figUa}).

These functions are fully normalizable for $n \in N$ since the hypergeometric functions
reduce to a polynomial of degree $n$ in $\cos^2z$.

The corresponding analytical expression for both even and odd-indexed mass levels
can be combined in a single expression
\beq m_{A}^2=l(4\alpha+ l), \ \ \ l=0,\ 1,\ 2, \dots \ . \label{positive alfa spectrum}\eeq
for $l$ even and odd respectively. It is worth noting that this result exhibits an exact
quantization condition on the massive modes of tensor and vector gauge
fields inherited from the extra dimension.

In the original $y$ variable,  these sets of solutions read
\beq u_{2n}(y;\, \alpha)= a_{2n}(\alpha)\,\, {}_2F_1\left(-n, 2\alpha+n, 2\alpha+\frac 1 2;
{\rm sech}^2y\right),\label{symmetric states}\eeq
and
\beq u_{2n+1}(y;\, \alpha)= a_{2n+1}(\alpha)(\tanh y) \,\, {}_2F_1\left(-n, 2\alpha+n+1,
2\alpha+\frac 1 2; {\rm sech}^2y\right),\label{antisymmetric states}\eeq
respectively, and are both finite polynomials in ${\rm sech}^2y$; $ a_{2n}(\alpha)$ and
$a_{2n+1}(\alpha)$ are normalization constants.

Since eq.(\ref{schrodinger}) involves an hermitian operator, for each value of $\alpha$
the spectrum is real and the above set of solutions is complete.


Now, regarding the second of eqs.(\ref{massive modes}), the procedure for the tensor massive modes
can be performed on the same footing. By means of the transformation
\beq \frac{dz}{dy}= e^{-\beta_2 \Lambda},\ \ \  w(y)=e^{-\alpha_2 \Lambda/2}W(z)
\label{change variables2}\eeq
we obtain similar, though not identical, modes for the Kalb-Ramond field.
%
%
\smallskip
\begin{figure} [ht]
\includegraphics{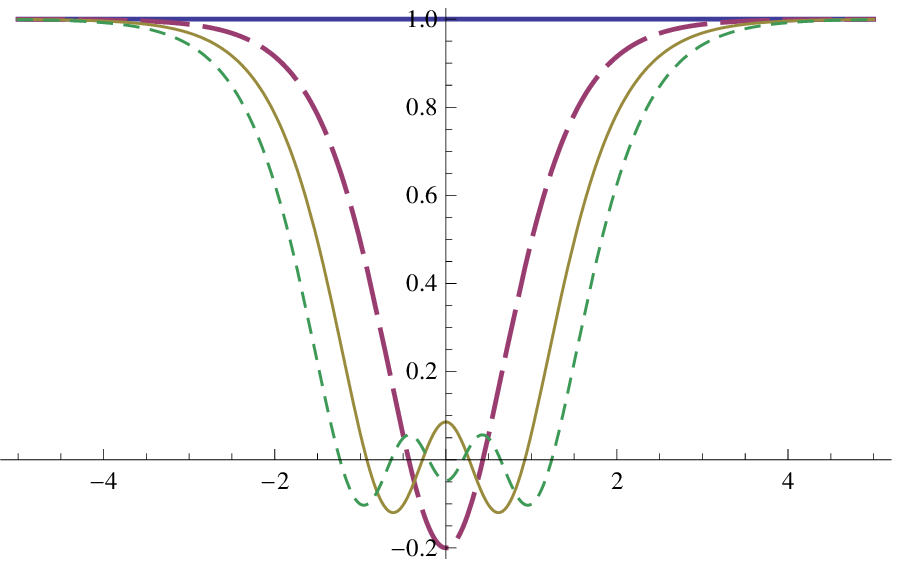}
\caption{Symmetric solutions $u_{2n}(y)$ for $\alpha=1$ and $n=0$ (blue solid line),
$n=1$ (red dashed line), $n=2$ (yellow thin solid line), $n=3$ (green short-dashed line); $n=0$
corresponds to the massless mode.}
\includegraphics{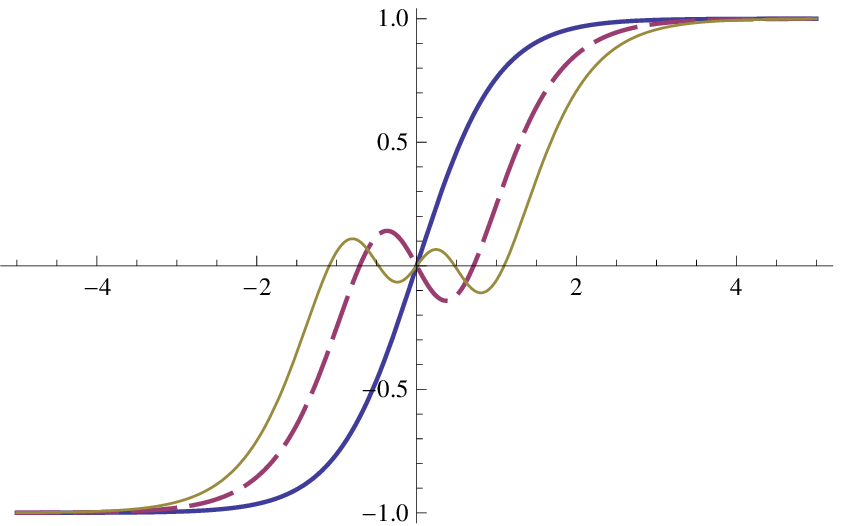}
\caption{Antisymmetric solutions $u_{2n+1}(y)$ for $\alpha=1$ and $n=0$ (blue solid line),
$n=1$ (red dashed line), $n=2$ (yellow thin solid line).}
\end{figure}
The difference results from the change $\alpha\rightarrow\alpha_2 = 4+\lambda\sqrt{3M^3}$
which modifies the spectrum and the eigenstates. This can be directly seen from the
set of equations above
by replacing $\alpha$ by $\alpha_2$. We include them for the sake of completeness
\beq w_{2n}(y;\, \alpha_2)= b_{2n}(\alpha_2)\,\, {}_2F_1\left(-n, 2\alpha_2+n, 2\alpha_2
+\frac 1 2; {\rm sech}^2y\right),\label{symmetric states w}\eeq
and
\beq w_{2n+1}(y;\, \alpha_2)= b_{2n+1}(\alpha_2)(\tanh y) \,\,
{}_2F_1\left(-n, 2\alpha_2+n+1, 2\alpha_2+\frac 1 2; {\rm sech}^2y\right),
\label{antisymmetric states w}\eeq
where $a_n(\alpha_2)$ and $b_n(\alpha_2)$ are the normalization constants of the KR modes,
and the corresponding spectrum is
\beq m_{B}^2=l(4\alpha_2+ l), \ \ \ l=0,\ 1,\ 2, \dots \ . \eeq
for $l$ even and odd respectively. 

\section{The complete analysis \label{sect complete}}

\begin{figure}
\includegraphics{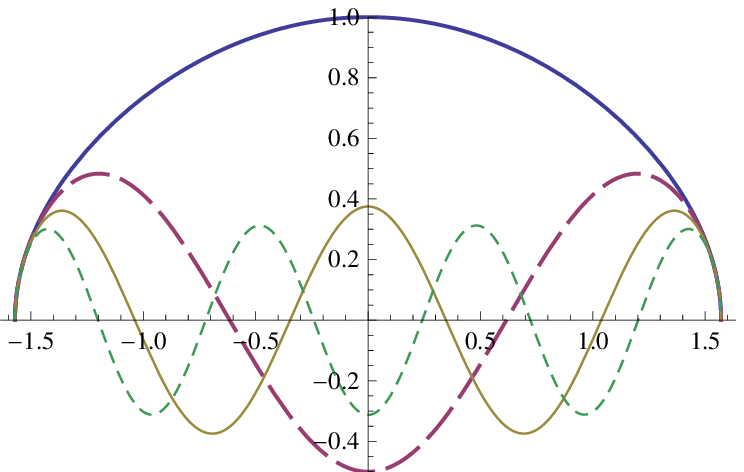}
\caption{\label{fig.solU(alpha_1_4_sim)} Symmetric $U(z)$ type functions for $\alpha = 1/4 $ and
$n=0$ (blue solid line), $n=1$ (red dashed line), $n=2$ (yellow thin-solid line),
$n=3$ (green short-dashed line);
$n=0$  is for the massless mode.} \label{fig.solU(alpha_1_4_sim)}
\includegraphics{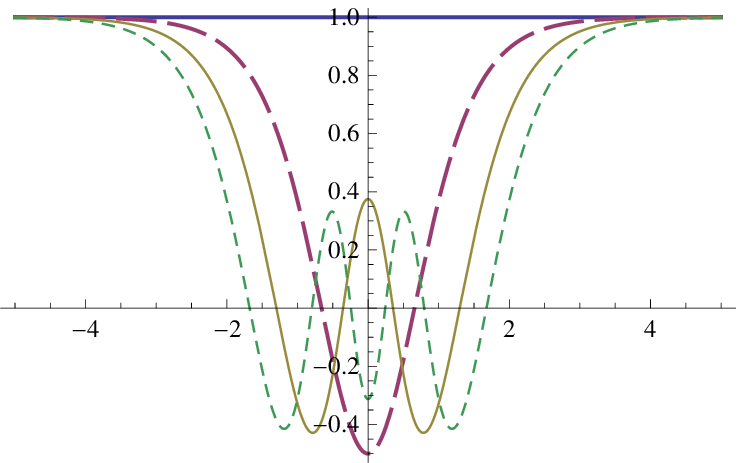}
\caption{\label{fig.sol1(alpha_1_4_sim)} Symmetric $u^{(1)}_{\rm even}(y)$ solutions for $\alpha =
1/4$ and $n=0$ (blue solid line), $n=1$ (red dashed line), $n=2$ (yellow thin-solid line),
$n=3$ (green short-dashed line); the constant corresponds to the massless gauge mode 5D factor
($n=0$, solid blue).}
\end{figure}

\begin{figure}
\includegraphics{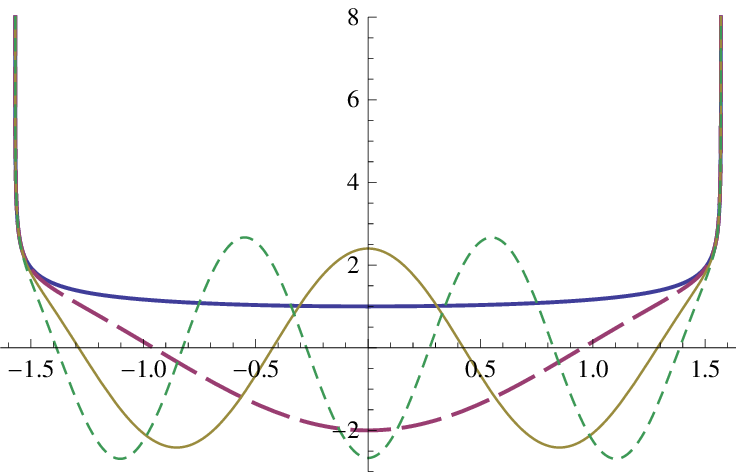}
\caption{\label{fig.solU(alpha_neg_1_8_sim)} Symmetric $U(z)$ functions for $\alpha = -1/8 $ and
$n=0$ (blue solid), $n=1$ (red dashed), $n=2$ (yellow thin-solid), $n=3$ (green short-dashed).
Note that all of them are divergent at $z=\pm \pi/2$.}
\includegraphics{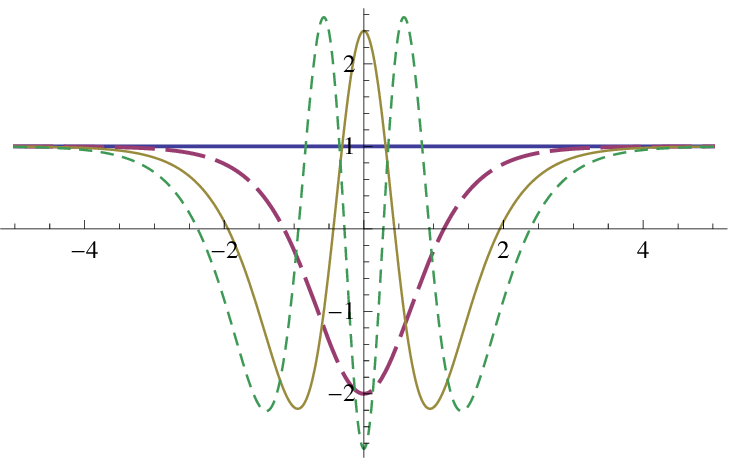}
\caption{\label{fig.sol1(alpha_neg_1_8_sim)} Symmetric $u^{(1)}_{\rm even}(y)$  solutions for
$\alpha = -1/8$ and
$n=0$ (blue solid), $n=1$ (red dashed), $n=2$ (yellow thin-solid), $n=3$ (green short-dashed).
 All are fully convergent in the $y$ space despite the
divergences in the $z$ coordinate. The constant corresponds to $n=0$ (solid-blue).}
\end{figure}

There is an important question to be clarified regarding the solution of our problem.
Often in the literature, the dimensional reduction procedure leads to the
above situation in which the original
differential equation for the quantum fields is mapped onto a Schrodinger equation by means of a
transformation like (\ref{change variables}). The calculation is then performed but not rarely
without due care
of the misleading non-relativistic quantum-mechanical aspect of the problem at hand.
Indeed,  following \cite{taseli}, in the last section we have limited our analysis
to solutions that are constrained by
the quantum-mechanical features of eq.(\ref{schrodinger}) and (\ref{potential tan2}).
However,  although the differential equation for $U$ is given by a Schrodinger operator,
 in the present context
$[-\frac{d^2}{dz^2}+ \mathfrak{V}(z)]$  is not of true hamiltonian nature,
as it is the case in actual Quantum Mechanics. This implies that a quantum-mechanical
reasoning may be physically incomplete, or even wrong in several aspects.
In fact, in the real transverse physical space the equation
of motion is (\ref{massive modes}) which is not Schrodinger at all.
Furthermore, for any $\alpha< 0$ it can be verified that the $m^2=0$ eigenvalue would not even be
allowed by the potential because the corresponding 'energy' ($E=2\alpha$)
is below its minimum. It would be in principle a wrong restriction since the zero-modes of
eq.(\ref{massive modes}) do exist and are simply given by $u$=cons, which is consistent
with $U=A\cos^{2\alpha}z$ for
any value of $\alpha \in R$ provided the hypergeometric functions are well defined
(analogously for $w$=cons and $W=B\cos^{2\alpha_2}z$).
Thus, the hamiltonian criterion is not justified since eq. (\ref{schrodinger}) does not
describe any
non-relativistic quantum particle.
Therefore, the existence and localization of a Maxwell zero-mode is possible in a wider
range than
just $\alpha\geq 1/2$ as implicitly assumed in the solutions found in the previous
Section (see e.g.
 Figs. \ref{fig.solU(alpha_1_4_sim)},\ref{fig.sol1(alpha_1_4_sim)} and
 Figs. \ref{fig.solU(alpha_neg_1_8_sim)},\ref{fig.sol1(alpha_neg_1_8_sim)}).
In order to constrain the values of the coupling constant
we shall take into account the analyticity of the solutions in the $y$ space,
and other physical reasons such as localization of zero-modes and the absence
of tachyons in the theory. The Kalb-Ramond field will
be analyzed as well by means of $\alpha_2$.

Let us emphasize that
%
the spectrum given by eq.(\ref{symmetric states}) and (\ref{antisymmetric states}) is complete
provided one imposes boundary conditions consistent with the hamiltonian problem,
namely, vanishing solutions $U(z)$ at $z=\pm\pi/2$ (in the $z$-space). Since
it is not enough to fully describe what is going on in the $y$-space, we need
to release these boundary conditions. Actually, in order to study all the relevant solutions of the
Sturm-Liouville problem (\ref{schrodinger}) related to the original differential equation, we should admit all the boundary conditions compatible
with finiteness of the $u, w$ functions in the $y$ space. For this, we should first relax nullification
at $z=\pm\pi/2$ and just require convergence in the open interval $(-\pi/2,\pi/2)$
admitting divergencies at the end points provided we get finite values after mapping back to the $y$ space.
This can be done by completely relaxing the parameters in the Schrodinger equation
above irrespective of quantum interpretations in the $z$-space. We will thus just
focus on $u(y)$ solutions which are the actual and direct 5D factors of physical gauge fields.
\smallskip
%

\subsection{Full gauge-field spectrum \label{subsect full}}

To start, note that in Section \ref{subsect exact} the so-called second solution
of the differential  equation has been just disregarded because it does not
vanish at $z=\pm\pi/2$ as expected for quantum mechanical bound-states.
Now, as a result of our discussion above, we will hereafter include this kind of solutions

%
%
\subsubsection{Symmetric eigenstates}
As a matter of fact, in the case of even states the complete set of
eigenfunctions to be considered is not just given by eq.(\ref{symmetric states}) but by both
\beq u^{(1)}_{\rm even}(y;\, \alpha, n)= a^{(1)}_{\rm even}(\alpha,n)\,\,
{}_2F_1\left(-n, 2\alpha+n, 2\alpha+\frac 1 2; {\rm sech}^2y\right),\label{symmetric states 1}\eeq
and
\beq u^{(2)}_{\rm even}(y;\, \alpha, n)= a^{(2)}_{\rm even}(\alpha,n)\,
({\rm sech}y)^{1-4\alpha}\,
{}_2F_1\left(\frac 1 2+n, -2\alpha+\frac 1 2-n, -2\alpha+\frac 3 2; {\rm sech}^2y\right),
\label{symmetric states 2}\eeq
in order to cover  the whole space of solutions of the differential operator.
Now, notice that nothing prevents $n$ from being a full real number
independently of $\alpha$, which is also in $R$. The only restriction
concerns the points where the Gauss functions are not well defined, namely
$\alpha =-\frac 1 4, -\frac 3 4, -\frac 5 4, \dots$ and $\alpha = \frac 3 4,
\frac 5 4, \frac 7 4, \dots$, for the first and second functions respectively.
In fact, in the second case $\alpha>1/4$ is already excluded due to the
$({\rm sech}y)^{1-4\alpha}$ factor and, in both, $\alpha\leq -1/4$ is
unimportant for we are not interested in
values which do not admit a localizable zero-mode.

Still, we can only accept continuous
and differentiable gauge fields,  so we have to analyze their behavior at the origin
(Gauss hypergeometric functions are convergent at any point else).
In case of eq.(\ref{symmetric states 1}), right and left derivatives at $y=0$ are
\beq \frac{du^{(1)}_{\rm even}}{dy}|_{y=0^\pm}=\pm \frac{\sqrt{\pi}\,4n(n+2\alpha)\,\Gamma(3/2+2\alpha)}
{(1+4\alpha)\,\Gamma(1-n)\,\Gamma(1+n+2\alpha)}.\eeq
Thus, derivatives are continuous  only if
\beq n=0, 1, 2, \dots,\ \ \ \ \ \ {\rm or}\ \ \ \ \ \ n+2\alpha =\ 0, -1, -2, \dots .\eeq
It shows that we can obtain well-behaved $u^{(1)}_{\rm even}$ solutions
(and localizable zero modes) not only for
 $n\in N$ but also for $n\in R$ provided it is related to $\alpha$ by
$n=-k-2\alpha,\ k\in N$, with $\alpha \in R - \{\alpha\leq-1/4\}$.
In fact, solutions $u^{(1)}_{\rm even}(y)$ are
symmetric under the change $n+2\alpha  \leftrightarrows -n$. Thus, in any such
cases solutions will be identical.
\smallskip
\begin{figure}
\includegraphics[width=7.5cm,height=5.5cm]{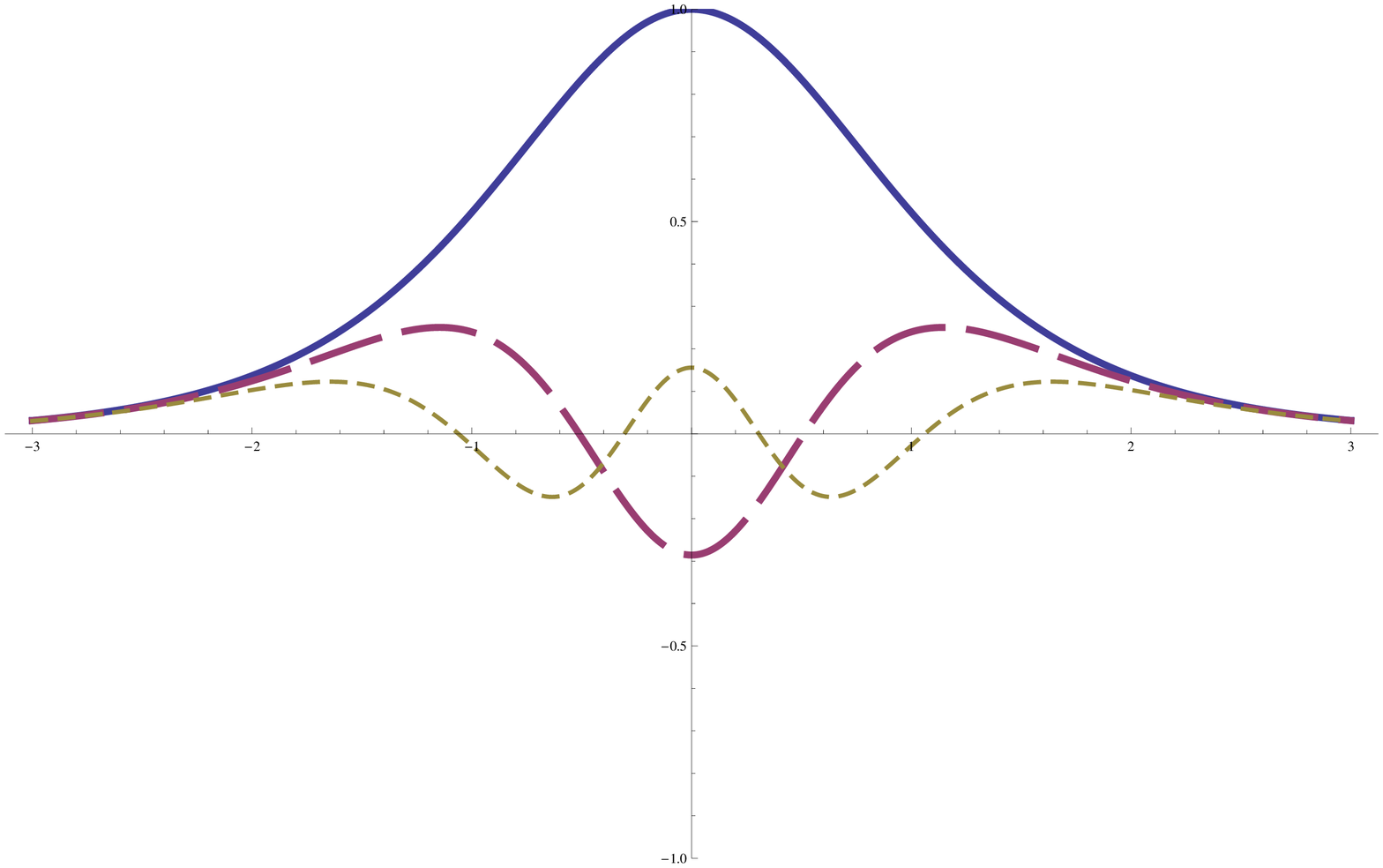}
\caption {\label{fig.sol2(alpha_neg 1_8_sim)} Second-symmetric solutions $u^{(2)}_{\rm even}(y)$
for $\alpha =  -1/8 $ and $n=-1/2-p$ where $p=0$ (blue solid), $p=1$ (red dashed), $p=2$ (yellow short-dashed).}
\end{figure}
\begin{figure}
\includegraphics[width=7.5cm,height=5.5cm]{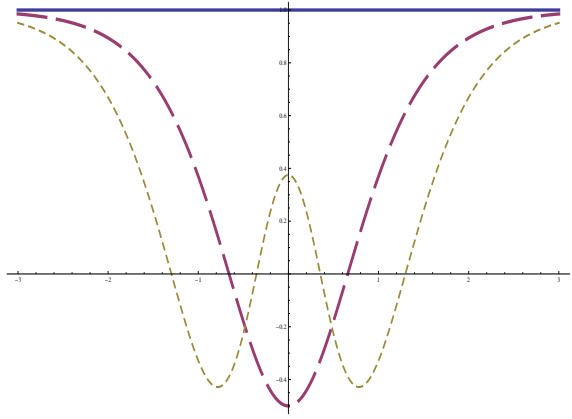}
\caption {\label{fig.sol2(alpha_1_4_sim)} Second-symmetric solutions $u^{(2)}_{\rm even}(y)$  for
$\alpha = 1/4  $ and $n=-1/2-p$ where $p=0$ (blue solid), $p=1$ (red dashed), $p=2$ (yellow short-dashed). Note the presence of a zero-mode for $p=0$
(blue solid line).}
\end{figure}

In case of eq.(\ref{symmetric states 2}), right and left derivatives at $y=0$ are
\beq \frac{du^{(2)}_{\rm even}}{dy}|_{y=0^\pm}=\pm
\frac{\sqrt{\pi}\,(4n^2+8n\alpha+4\alpha-1)\,\Gamma(5/2-2\alpha)}
{(4\alpha-3)\,\Gamma(3/2+n)\,\Gamma(3/2-n-2\alpha)}.\eeq
These  derivatives are continuous provided
\beq n= -1/2, -3/2, \dots,\ \ \ \ \ \ {\rm or}\ \ \ \ \ \ n+2\alpha = 1/2, 3/2, \dots \eeq
and we can also have well-behaved $u^{(2)}_{\rm even}$ solutions
(and localizable zero modes) for $n$ real when
$n=-k/2\ {\rm or}\ n= k/2 - 2\alpha,\ k\in N_0$, with $\alpha \in (-1/4<\alpha\leq 1/4]$.
See Figs. \ref{fig.sol2(alpha_neg 1_8_sim)},\ref{fig.sol2(alpha_1_4_sim)}.
Solutions $u^{(2)}_{\rm even}(y)$ are also symmetric under the change
$n+2\alpha  \leftrightarrows -n$.

In order to identify symmetric zero-modes, we analyze the even spectrum,
which is given by
\beq m^2=4n(n+2\alpha).\label{even spectrum}\eeq
From this we can pick the following possibilities: $u^{(1)}_{\rm even}(n=0, \alpha>-1/4)$, and
$u^{(1)}_{\rm even}(n=-2\alpha, \alpha>-1/4)$, and $u^{(2)}_{\rm even}(n=0, \alpha=1/4)$ and
$u^{(2)}_{\rm even}(n=-1/2, \alpha=1/4)$ where analyticity together with the localization
restriction have been assumed. Note that these are in fact constants for every value of $\alpha$,
as already pointed out.

If we now look for a massive spectrum free of tachyons, other constraints will restrict the
space of legal solutions. Since the even spectrum is given by eq.(\ref{even spectrum}),
tachyons are avoided only in the following cases
\bea && (n=1, \alpha>-1/2),\ (n=2, \alpha> -1), \dots \\
&& (n=-1-2\alpha, \alpha> -1/2),\ (n=-2-2\alpha, \alpha> -1),\dots\nonumber \eea
for $u^{(1)}_{\rm even}$, and
\bea
&& (n=-1/2, \alpha< 1/4),\ (n=-3/2, \alpha<3/4),\ (n=-5/2, \alpha\leq5/4),\dots \\
&& (n=1/2-2\alpha, \alpha<1/4),\ (n=3/2-2\alpha, \alpha<3/4),\ (n=5/2-2\alpha, \alpha<5/4),\dots
\nonumber \eea
for $u^{(2)}_{\rm even}$. In each case, these additional constraints have to be considered
on top of the corresponding analyticity restrictions shown above.
This demonstrates that there is no risk of tachyons among analytic symmetric eigenstates
when localized zero modes are demanded in the theory. There is however an important remark
to be done in this respect when just one of the gauge fields is demanded to deposit zero modes
on the brane. We will come to this point in   Section \ref{subsect phases}.

\subsubsection{Antisymmetric eigenstates}
\smallskip
\begin{figure}
\includegraphics[width=7.5cm,height=5.5cm]{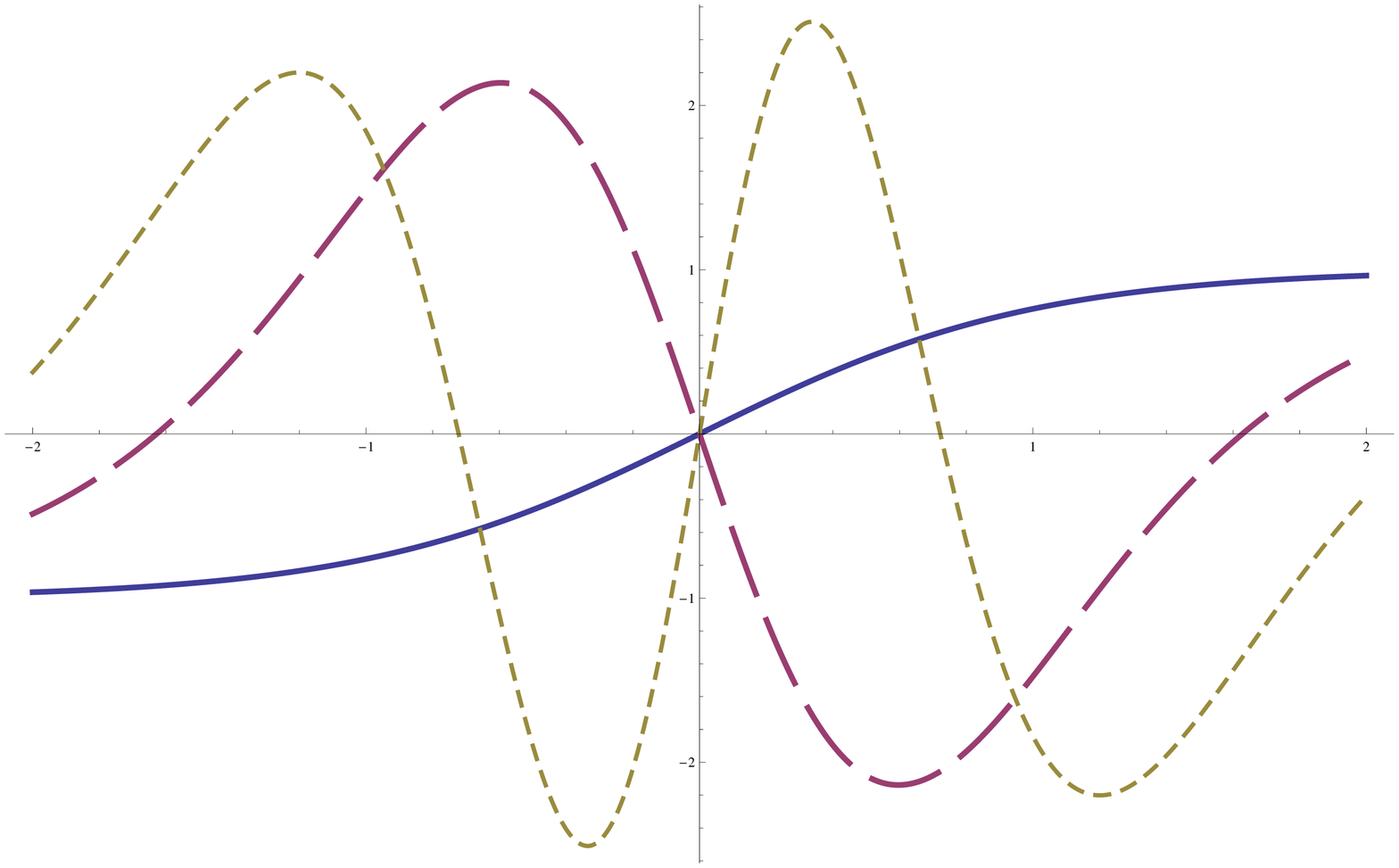}
\caption {\label{fig.sol1(alpha_neg 1_8_Asim)} Antisymmetric solution $u^{(1)}_{\rm AS}(y)$ for
$\alpha = -1/8 $ and $n=0$ (blue solid), $n=1$ (red dashed), $n=2$ (yellow short-dashed line).}
\end{figure}
\begin{figure}
\includegraphics[width=7.5cm,height=5.5cm]{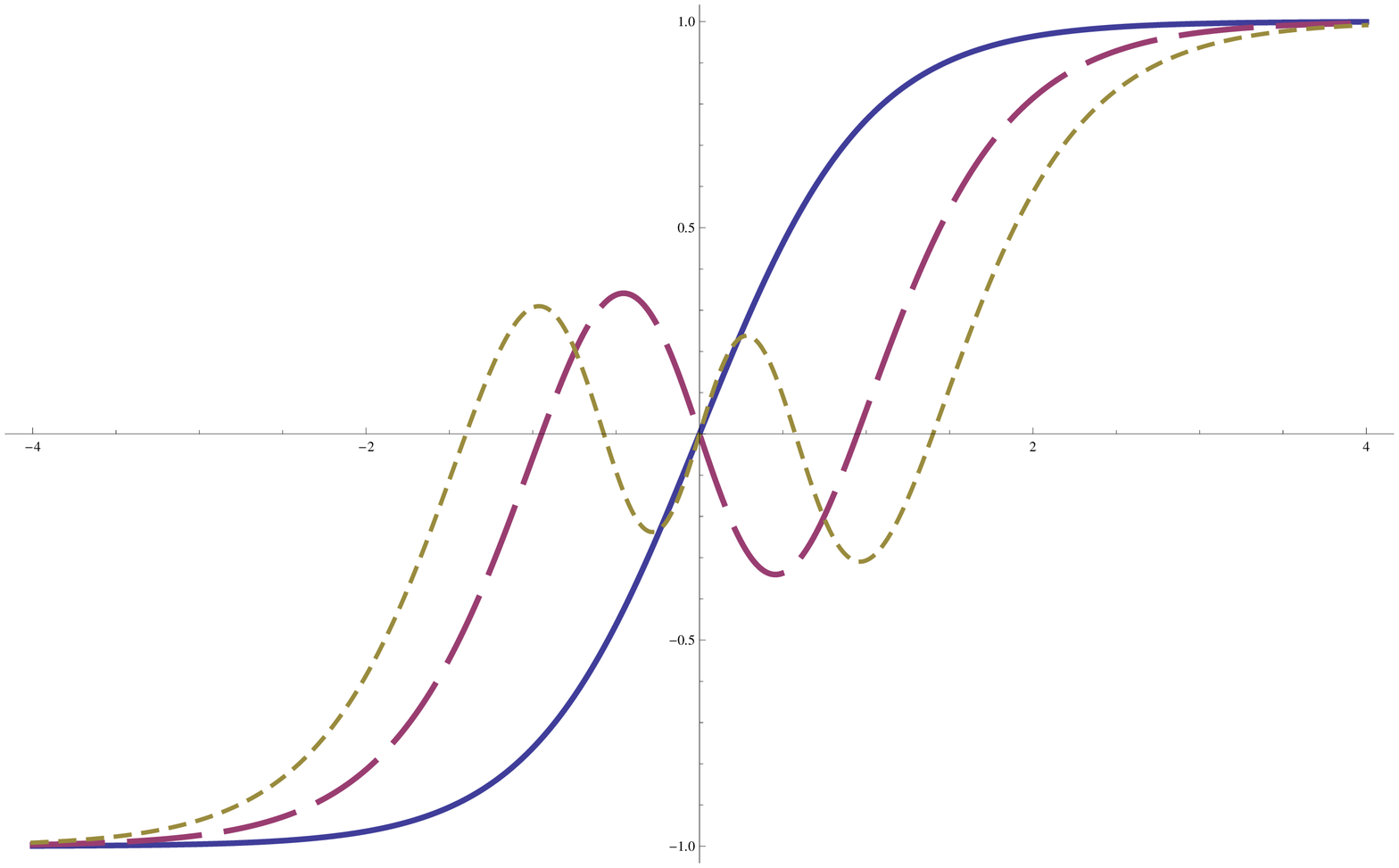}
\caption {\label{fig.sol1(alpha_3_8_Asim)} Antisymmetric solution $u^{(1)}_{\rm AS}(y)$  for
$\alpha = 3/8 $ and $n=0$ (blue solid), $n=1$ (red dashed), $n=2$ (yellow short dashed line).}
\end{figure}

Regarding antisymmetric eigenfunctions, the complete set of solutions is
\beq u^{(1)}_{\rm AS}(y;\, \alpha,n)= a^{(1)}_{\rm AS}(\alpha,n)(\tanh y) \,\,
{}_2F_1\left(-n, 2\alpha+n+1, 2\alpha+\frac 1 2; {\rm sech}^2y\right),\label{antisymmetric states 1}\eeq
and
\bea u^{(2)}_{\rm AS}(y;\, \alpha,n)= && a^{(2)}_{\rm AS}(\alpha,n)\,({\rm sech}y)^{1-4\alpha}
\,(\tanh y) \nonumber\\
&&{}_2F_1\left(1/2+n, -2\alpha-n+\frac 1 2, -2\alpha+\frac 3 2; {\rm sech}^2y\right).
\label{antisymmetric states 2}\eea
The first of these is not defined for $\alpha =-\frac 1 4, -\frac 3 4, -\frac 5 4, \dots$
and the second for $\alpha = \frac 3 4,  \frac 5 4, \frac 7 4, \dots$ . As before, in the
second case $\alpha>1/4$ is excluded due to the divergency of $({\rm sech}y)^{1-4\alpha}\tanh y$,
while $\alpha\leq -1/4$ is unimportant in both cases for we will not be interested in values
which do not admit localizable zero-modes.
%

\smallskip
\begin{figure}
\includegraphics[width=7.5cm,height=5.5cm]{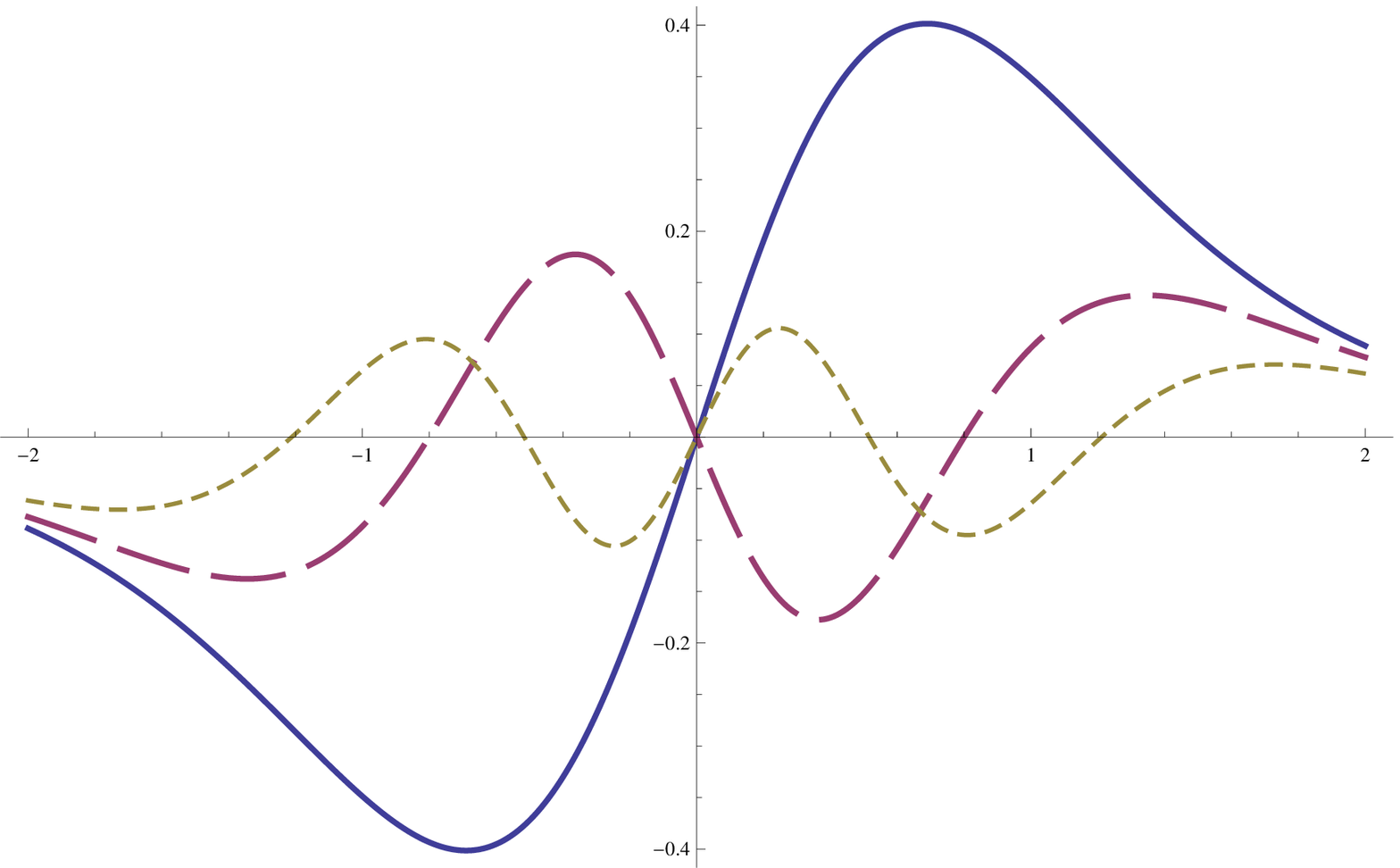}
\caption {\label{fig.sol2(alpha_neg 1_5_sim)} Second-antisymmetric solution $u^{(2)}_{\rm AS}(y)$
for $\alpha = -1/5 $  and $n=-3/2-p$ where $p=0$ (blue solid), $p=1$ (red dashed), $p=2$ (yellow short-dashed line).}
\end{figure}
\smallskip
\begin{figure}
\includegraphics[width=7.5cm,height=5.5cm]{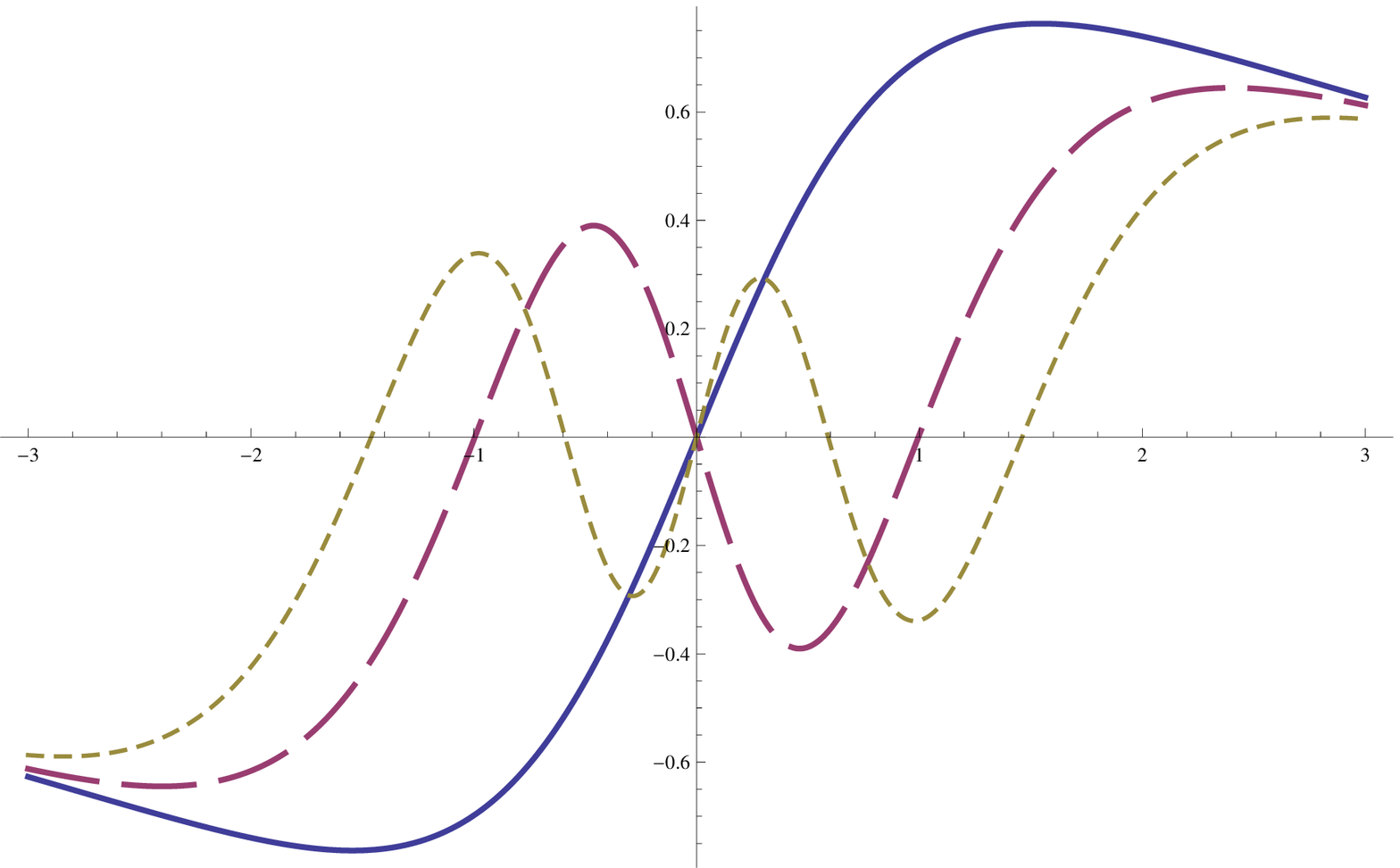}
\caption {\label{fig.sol2(alpha_1_5_sim)} Second-antisymmetric solution $u^{(2)}_{\rm AS}(y)$  for
$\alpha = 1/5 $  and $n=-3/2-p$ where $p=0$ (blue solid), $p=1$ (red dashed), $p=2$ (yellow short-dashed line). }
\end{figure}

Unlike even functions, antisymmetric functions can be discontinuous at the origin. Actually, we have
\beq u^{(1)}_{\rm AS}(y=0^{\pm})=\pm \ \frac{\sqrt{\pi}\,\Gamma(1/2+2\alpha)}
{\Gamma(-n)\,\Gamma(1+n+2\alpha)}\eeq
although left and right derivatives coincide
\beq \frac{du^{(1)}_{\rm AS}}{dy}|_{y=0^\pm}= {}_2F_1\left(-n, 2\alpha+n+1, 2\alpha
+\frac 1 2; 1\right).\eeq
Thus, in order to avoid discontinuities of the gauge field we require
\beq n= 0, 1, 2, \dots,\ \ \ \ \ \ {\rm or}\ \ \ \ \ \ n+2\alpha = -1, -2, \dots\ .\eeq
and $\alpha\neq-\frac 14$ which is already forbidden.

The second antisymmetric solution can also be discontinuous depending on the value
of the parameters.
The analytical right and left expressions are
\beq u^{(2)}_{\rm AS}(y=0^{\pm})= \pm\frac{\sqrt{\pi}\,\Gamma(3/2-2\alpha)}{\Gamma(3/2+n)
\,\Gamma(1/2-n-2\alpha)},\eeq
so that these eigenfunctions are continuous only for
\beq n= -3/2, -5/2, \dots,\ \ \ \ \ \ {\rm or}\ \ \ \ \ \ n+2\alpha = 1/2, 3/2, \dots\ \eeq
and $\alpha \neq  3/4, 5/4, \dots $, all of which are  above the region of convergence
of the hyperbolic overall factor.
The exact derivatives of this family of eigenstates are
\beq \frac{du^{(2)}_{\rm AS}}{dy}|_{y=0^\pm}= {}_2F_1\left(3/2+n, -2\alpha-n+\frac 1 2,
-2\alpha+\frac 3 2; 1\right)\eeq
which are well defined for all $\alpha\neq\frac 34, \frac 54,\dots$ (these are all
 bigger than 1/4
and thus outside the range of convergence of $u^{(2)}_{\rm AS}(y;\, \alpha,n)$ as
already signaled).

In the antisymmetric case, the mass spectrum is given by
\beq m^2=(2n+1)(2n+1+4\alpha)\eeq
so zero-modes are not possible considering the analyticity restrictions of these eigenfunctions.

In order to avoid tachyons, besides conditions for existence and localization,
in the case of $u^{(1)}_{\rm AS}$ we shall require
\bea && (n=0, \alpha> -1/4),\ (n=1, \alpha>-3/4),\ (n=2, \alpha> -5/4),\ \dots \nonumber\\
&& (n=-1-2\alpha, \alpha>-1/4),\ (n=-2-2\alpha, \alpha>-3/4),\ \dots\ \eea
while, regarding $u^{(2)}_{\rm AS}$,  the non-tachyon constraints are
\bea && (n=-3/2, \alpha<1/2),\ (n=-5/2, \alpha<1),\  \dots \nonumber\\
&& (n=1/2-2\alpha, \alpha ),\ (n=3/2-2\alpha, \alpha),\ \dots\ .\eea
Tachyons are again impossible considering the analyticity constraints of the functions
and the restrictions on $\alpha$ due to localization of zero modes.
~~

All this analysis can be straightforwardly repeated for the Kalb-Ramond field just by changing
 the parameter $\alpha \rightarrow \alpha_2$.

\subsection{Phase interpretation picture \label{subsect phases}}

The present field theory can be classified into different phase configurations
depending on the value of the dilaton coupling constant.
Since the exact value of this coupling constant should be specified by an ascendant theory
we will again consider the whole range for this analysis.
In order to characterize the possible different
phases in the theory we can use the shape of the analog potential functions (\ref{potential tan2})
appearing in eq.(\ref{schrodinger}).

\begin{figure}
\includegraphics[width=7.5cm,height=4.5cm]{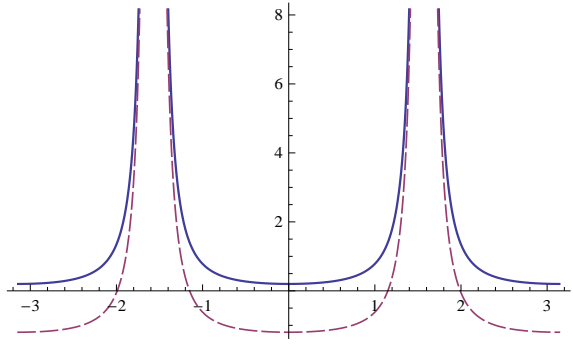}
\caption {\label{fig.pot up} $\mathfrak{V}_A(z)$ for $\alpha =
-1/10$ (solid blue) and $\alpha =
1/2+1/10$  (red dashed).}
\end{figure}

\begin{figure}
\includegraphics[width=7.5cm,height=4.5cm]{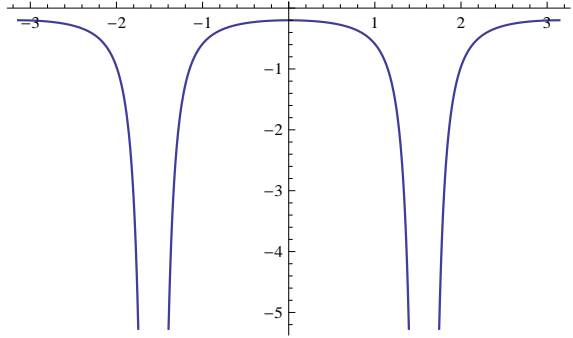}
\caption {\label{fig.pot down}$\mathfrak{V}_A(z)$ for $\alpha = 1/2-1/10$.}
\end{figure}

In the parameter region $\alpha\geq 1/2$ or $\alpha\leq0$, eq.(\ref{potential tan2})
defines a bounding analog potential $\mathfrak{V}(z) \geq 0$ (see Fig. \ref{fig.pot up}).
On the contrary, in the complementary region of the $\alpha$ parameter, namely $0<\alpha<1/2$,
$\mathfrak{V}(z)$ is negative allowing for a different physical picture (see Fig. \ref{fig.pot down}).
Even in this case, when the mass is zero we get as announced
a constant value for the $u$ solution in the $y$ space (note that this legal mode is forbidden
in a strictly hamiltonian interpretation of the problem).

If $\alpha\leq-1/4$, then $\lambda\leq\lambda_0$
and a localized gauge zero-mode is impossible as pointed out earlier. However,
if $\alpha_2\leq-1/4$ the coupling is $\lambda\leq\lambda_0/2$ and then,
although no Kalb-Ramond zero-mode is localized,  there can be a Maxwell
zero-mode deposited on the membrane. If $0<\alpha<1/2$, then
$\frac{16}{17}\lambda_0<\lambda<\frac{14}{17}\lambda_0$ which lies
also between $\lambda_0<\lambda<\lambda_0/2$ and again no Kalb-Ramond
but just a Maxwell zero-mode has a (finite) contribution to the effective action.
If $0<\alpha_2<1/2$,
then both zero-modes should be localizable.
Note that, if the energy criterion was valid, a Maxwell zero-mode would
only exist for $\lambda>\frac{16}{17}\lambda_0$ and, on the same footing
a localized zero-mode for the Kalb-Ramond field would only exist
provided $\alpha_2> 0$, namely $\lambda>\frac{8}{17}\lambda_0$.

In terms of the dilaton coupling constant, the simultaneous analysis of
both gauge fields results in the following conclusion: a gauge zero mode
exists and is localizable provided $\lambda>\lambda_0$  with the undermentioned details.
\textit{i}) In  $\frac{16}{17}\lambda_0<\lambda<\frac{14}{17}\lambda_0$ the theory possess
a localized Maxwell zero-mode
and a tower of normalizable massive Maxwell modes resulting from an analog potential $\mathfrak{V}_A(z)>0$,
together with normalizable Kalb-Ramond massive states related to an analog potential $\mathfrak{V}_B(z)<0$.
\textit{ii}) In  $\frac{14}{17}\lambda_0<\lambda<\frac 12\lambda_0$ there is a Maxwell zero-mode
and normalizable massive Maxwell modes related to $\mathfrak{V}_A(z)<0$
together with normalizable Kalb-Ramond massive states related to $\mathfrak{V}_B(z)<0$.
\textit{iii}) In  $\frac 12\lambda_0<\lambda<\frac{8}{17}\lambda_0$ a Kalb-Ramond zero-mode is also
deposited on the previous configuration.
\textit{iv}) In  $\frac{8}{17}\lambda_0<\lambda<\frac{7}{17}\lambda_0$
the model presents a Maxwell zero-mode
and massive Maxwell modes as resulting from $\mathfrak{V}_A(z)>0$, together with
a Kalb-Ramond zero-mode
and massive Kalb-Ramond states coming from a potential $\mathfrak{V}_B(z)<0$.
\textit{v}) If $\lambda>\frac{7}{17}\lambda_0$, both Maxwell and Kalb-Ramond zero-modes are localized
and normalizable Kaluza-Klein eigenstates can be related to $\mathfrak{V}_{A, B}(z)>0$.

We thus conclude that depending on the value of the dilaton coupling
the theory sits in one among several phases,
which are given by the global sign of the analog potentials defined in transverse space.
The two potentials change dramatically as $\alpha$ (or $\alpha_2$)
take values on one side or another of 0 and 1/2. Thus,
from $\lambda=\frac{16}{17}\lambda_0-\epsilon$ to $\lambda=\frac{16}{17}\lambda_0+\epsilon$,
the analog potential  flips from Fig.
\ref{fig.pot up} (solid) to Fig. \ref{fig.pot down}, and for
$\lambda=\frac{14}{17}\lambda_0+\epsilon$ instead of $\frac{14}{17}\lambda_0-\epsilon$
the analog potential is that of Fig. \ref{fig.pot up} (dashed) instead of
Fig. \ref{fig.pot down}.
Analogously,
from $\frac{8}{17}\lambda_0+\epsilon$ to $\frac{8}{17}\lambda_0-\epsilon$
the potential for KR modes
$\mathfrak{V}_B(z)$ would invert its concavity, but it would be back the same
for $\lambda$ above $\frac{7}{17}\lambda_0$.

\begin{figure}
\includegraphics{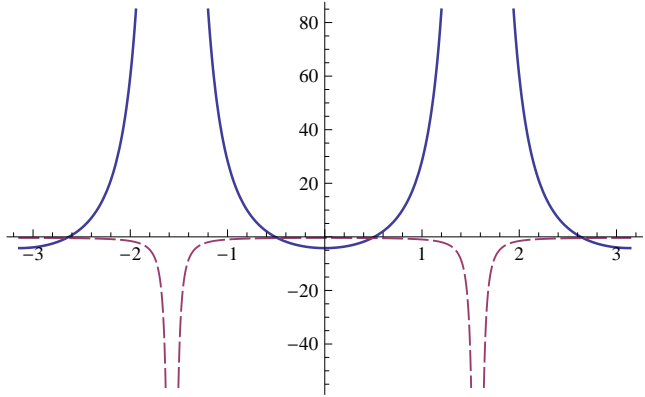}
\caption {\label{potencialesMQ}$\mathfrak{V}_A(z)$ (solid) and $\mathfrak{V}_B(z)$ (dashed) for
$\alpha_2=1/5$.  It corresponds to a dilaton coupling constant in
$\frac{8}{17}\lambda_0<\lambda<\frac{7}{17}\lambda_0$.}
\end{figure}

Particularly interesting is the interval
$\frac{8}{17}\lambda_0<\lambda<\frac{7}{17}\lambda_0$ where  Maxwell and Kalb-Ramond
fields experience totally different analog potentials at the same time
(see Fig. \ref{potencialesMQ}) and both have zero-modes localized on the membrane.
\smallskip

Finally, it is worth noting that if one just demands the theory to possess a localized Maxwell
zero-mode but no such condition is put on the Kalb-Ramond field, there is a full segment
of possible values of the dilaton coupling constant where several kinds of gauge messengers could
co-exist. Specifically,
in the interval going from $\lambda_0$ to $\lambda_0/2$, namely $\alpha\in(-1/4,15/8]$.
When $\lambda>\lambda_0$, we can write $\alpha=-1/4+\epsilon$ and $\alpha_2=-9/2+2\epsilon$.
In this cases Kaluza-Klein eigenstates are well defined for both fields but some Kalb-Ramond modes are
compatible with $m_B^2<0$. For example, when $\epsilon$ is sufficiently small,
the eigenfunctions $u^{(1)}_{even}(\alpha_2,n, y)$ for
$n= 5, 4, 3, 2, 1, -1-2\alpha_2, \dots,  -5-2\alpha_2$,
and the eigenfunctions $u^{(1)}_{AS}(\alpha_2,n, y)$ for
$n= 8, 7, \dots, 1, 0, -1-2\alpha_2, -2-2\alpha_2, \dots,  -9-2\alpha_2$
are all well-defined  tachyonic modes (for larger values of $\epsilon$
the number of tachyons is smaller).
Thus, if the dilaton coupling constant happens to be in $(\lambda_0, \lambda_0/2]$,
several tachyons would co-exist together with a tower of massive eigenstates of
both Maxwell and Kalb-Ramond fields and a Maxwell localized zero-mode.

{}

\subsection{Final remarks}

In order to determine that massive modes are suppressed as compared with zero-modes,
one can evaluate the variation of the effective gauge coupling as a function of the
Kaluza-Klein photonic masses.
%
\begin{figure}
\includegraphics{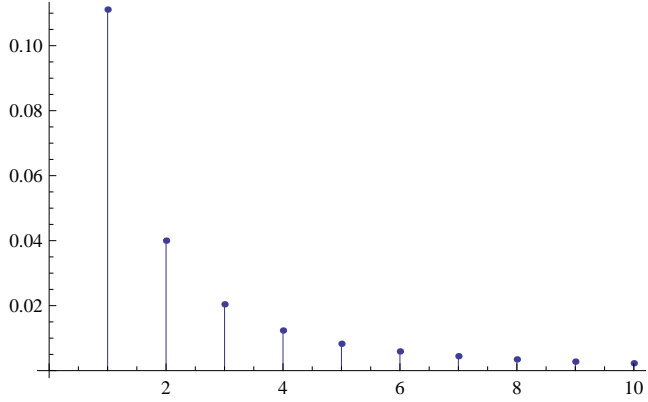}
\caption {\label{discrete1}Sequence of $(u^{(1)}_{\rm even}(0))^2$ values for
$n= 1, 2,\dots, 10$
displays  relative weights on the brane The $n=0$ mode is about one order bigger
than $n=1$ and is
not in the figure.}
\end{figure}
%
Since in the non-relativistic limit the coupling of massive modes with matter on
the brane develops a Yukawa type potential,
it is natural that massive contributions are strongly attenuated as compared
with the Coulomb potential.
To show that this quantity is a decreasing function of $m_A$ we should
evaluate the different coefficients
that multiply the  four-dimensional action
\beq\sim\int dy\ \left(u^2_{m_A=0}(y)+\sum_n\ u_{m_A(n)}^2(y)\right)\int d^4x f_{\mu\nu}f^{\mu\nu}.\eeq
In order to simplify this computation we shall assume that the coupling with the brane
takes place precisely on 4D ordinary space-time, namely at $y=0$. It is there where the
relevant dynamics should be much stronger. Thus, the effective 4D electrostatic potential would read
\beq V(r)\sim{q_1q_2}\left( \frac{A^2}{r}+\sum_{n} \frac{e^{-m_A(n)\ r}}{r} u_{m_A(n)}^2(0)\right)\eeq
\begin{figure}
\includegraphics{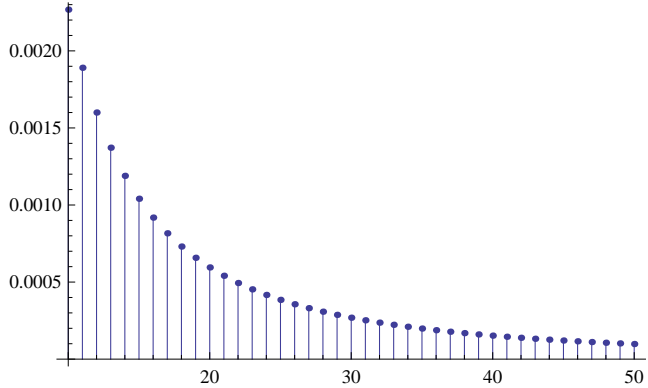}
\caption {\label{discrete2}Sequence of $(u^{(1)}_{\rm even}(0))^2$ values for $n= 10,\dots, 50$
displays relative weights on the brane ($\alpha=1/2$).}
\end{figure}
\noindent where $q_1, q_2$ are two test charges separated a distance $r$
in ordinary 3D space and the Kaluza-Klein eigenvalues $m_A$ grow with $|n|$
as we have seen. See e.g. Fig. \ref{modo high n} where the $u^{(1)}_{\rm even}(y)$
modes $n= 0, 2, 50$ are fully displayed.
By analytical calculation, for some values of $\alpha(\lambda)$ we can appreciate
that at the origin each contribution is indeed rapidly decreasing with mass.
See Figs. \ref{discrete1},\ref{discrete2}. This, together with the negative exponential
factor, essentially
decouples the massive modes from the physics on the domain wall.
Far from the membrane,
all massive modes become constants, as much as zero-modes are, and as a consequence
5D phenomenology results completely modified from ordinary 4D electromagnetism.
The same analysis can be performed for the Kalb-Ramond modes with the same result.
See e.g. Refs.\cite{RS, csaki al} for the study of this issue in the case of gravity.
\smallskip
\begin{figure}
\includegraphics{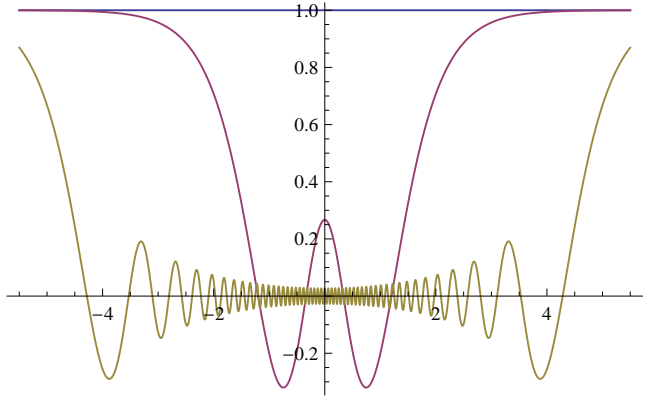}
\caption {\label{modo high n} Some $u^{(1)}_{\rm even}(y)$ modes $n= 0, 2, 50$ (horizontal blue line,
two-minima red line, highly oscillating yellow line)
exhibit their relative weights. Here $\alpha=3/8$.}
\end{figure}
{}
{}
\section{Conclusions \label{sect conclusion}}
In this paper we have analyzed the existence and localization of Maxwell and
Kalb-Ramond propagation  modes in a warped extra-dimensional universe.
As a result of gauge invariance, the common knowledge is that charged fields
must be confined into a four-dimensional brane-world while only neutral fields
(under standard model interactions) can be bulk degrees of freedom interacting with the brane.
Fermionic and gravitational fields have been
localized in the past rather easily but gauge fields have demanded more effort.
As pointed out in \cite{keha-tamva, youm1}, the presence of the dilaton
happens to be necessary for gauge field modes to be localized in brane-world theories.
To analyze this question in a relevant multivacua field theory, we have studied a sine-Gordon like
thick membrane that bounces at the (non-compact) extra-coordinate origin. We have focused
a five-dimensional metric with two warping functions and two interacting
scalar fields. One of them represents the membrane itself and the other
represents the dilaton in a field theoretic scenario. The corresponding
action includes a potential functional depending on both scalar fields,
which is a nontrivial deformation of a sine-Gordon potential and is dynamically consistent
with the gravity background. The solution to this action represents a
universal framework that interacts with two distinct five-dimensional gauge fields.

To the best of our knowledge, close expressions
for all the gauge modes in a 5D space-time have been obtained here for the first time,
particularly for the present combination of field degrees of freedom.
Regarding the propagation modes of the Kalb-Ramond field, so far only
simple qualitative calculations have been developed in this context. Here,
besides performing an analytical approach, we also included the electromagnetic field
to treat both of them simultaneously and exhibit their actual differences.

After a detailed analysis of the bulk equations of
motion of Kalb-Ramond and Maxwell fields, we have presented the full spectrum
of the gauge field problem. The exact variation of the whole set of eigenstates
and eigenvalues, with the dilaton coupling constant $\lambda$, was discussed in detail
and the exact dependence of vector and tensor gauge modes with the extra coordinate was analytically
computed in all the cases. As one of our results, we have proven that localization of a
Maxwell zero-mode on a smooth thick domain wall embedded in an five-dimensional
world is granted for a dilaton coupling constant above $\lambda_0$.

Depending on the value of $\lambda$,
different phases of the analog potentials related to the equations of motion
of Maxwell and Kalb-Ramond fields can arise.
If the dilaton coupling constant is on one side or another of determined particular values, the related
analog potentials can be dramatically different, allowing for an interpretation of
subjacent different physical scenarios (see Sect.\ref{subsect phases}). 
Regarding finite Kaluza-Klein modes,
we have analytically found the full spectra of Maxwell and Kalb-Ramond
fields and shown that massive modes are strongly suppressed on the brane
so that ordinary four-dimensional gauge interactions are not modified in
the present set up (see Sect.V C). Among this infinite tower of massive modes, a handful
of Kalb-Ramond tachyons could not be excluded unless the dilaton coupling
happens to be greater than $\lambda_0/2$. Notably, in this case the theory would lose
tachyons but would gain a localized Kalb-Ramond zero-mode. {In this way, the present model
has been useful to show that a localized zero-mode is not only possible for a 5D electromagnetic field
but also, and simultaneously, for a 5D Kalb-Ramond field, depending on just the 
value of the dilaton coupling constant.
As a general conclusion, the above results help showing that the model presented here
is an interesting arena to discuss extra-dimensional physics and that our ordinary 4D world
seems to be compatible with a higher dimensional universe, apparently of a stringy brane nature.}

\begin{acknowledgments}
H.R.C. acknowledges FUNCAP for financial support and F.A. Schaposnik for useful comments.
\end{acknowledgments}

\end{document}